\documentclass[prb,aps,showpacs,floats]{revtex4}
\usepackage{amssymb}
\usepackage{graphicx}
\begin{document}

\title{Low-temperature electrical resistivity in paramagnetic spinel  LiV$_2$O$_4$}

\author{V. Yushankhai$^{1,2}$, T. Takimoto $^{3}$, and  P. Thalmeier$^{4}$}

\affiliation{ $^1$Joint Institute for Nuclear Research, 141980 Dubna, Russia\\
$^2$Max-Planck-Institut f\"ur Physik Komplexer Systeme, N\"othnitzer Stra\ss e 38,
             D-01187 Dresden, Germany\\
 $^3$Asia Pacific Center for Theoretical Physics, Pohang University of Science and Technology,
      Pohang, Gyeongbuk 790-784, Korea\\
 $^4$Max-Planck-Institut f\"ur Chemische Physik fester Stoffe, D-01187 Dresden,
  Germany\\}

\date{\today}


\begin{abstract}

The 3$d$ electron spinel compound LiV$_2$O$_4$ exhibits heavy fermion
behaviour below 30K which is related to  antiferromagnetic spin
fluctuations strongly enhanced in an extended region of momentum
space. This mechanism explains enhanced thermodynamic quantities and
nearly critical NMR relaxation in the framework of the selfconsistent
renormalization (SCR) theory. Here we show that the low-$T$ Fermi
liquid behaviour of the resistivity and a deviation from this
behavior for higher $T$ may also be understood within that context.
We calculate the temperature dependence of the electrical resistivity
$\rho(T)$ assuming that two basic mechanisms of the quasiparticle
scattering, resulting from impurities and spin-fluctuations, operate
simultaneously at low temperature. The calculation is based on the
variational principle in the form of a perturbative series expansion
for $\rho(T)$. A peculiar behavior of $\rho(T)$ in LiV$_2$O$_4$ is
related to properties of low-energy spin fluctuations
whose $T$-dependence is obtained from SCR theory.\\

\end{abstract}

\pacs{71.27.+a, 71.10.-w,  72.10.Di,  72.15.-v}

\maketitle


\section{Introduction}
The metallic vanadium oxide  LiV$_2$O$_4$ has attracted much
attention after a heavy fermion behavior in this 3$d$-electron system
was discovered.\cite{Kondo97,Johnston00,Kondo99} The cubic spinel
LiV$_2$O$_4$ has the pyrochlore lattice of vanadium ions (in the
mixed valence state V$^{3.5+}$) and shows metallic conduction and no
long-range magnetic ordering for any measured temperatures at ambient
pressure. So far the origin of the heavy fermion quasiparticle
formation observed in this compound for $T<10$K remains to be a
controversial subject, however, effects of  electronic
correlations and the geometrical frustration of the pyrochlore
lattice are supposed to be key aspects of the problem.

The quasiparticle mass enhancement is expected when a metallic system
is driven by strong electron correlations to a vicinity of a charge
and/or spin phase transition at low $T$. In that case, the
charge/spin disordered ground state on the metallic side of the
transition in the strongly correlated system LiV$_2$O$_4$ is
sustained because a long-range order with a particular ordering
(critical) wave vector ${\bf Q}_c$ is prevented by the geometrical
frustration. Expressed differently, the system cannot choose a unique
wave vector of an ordered structure which minimizes the free energy.
Instead, it is frustrated between different structures with different
critical wave vectors ${\bf Q}_c$'s and equally low free energy. 
For instance, low energy spin fluctuations are expected to be present
in a very large region of momentum space which is the signature of
frustrated itinerant magnetism. This is in contrast to non-frustrated
systems where the fluctuations are confined to the immediate vicinity
of a unique incipient ordering vector.

This scenario for frustrated itinerant magnetism  was recently
investigated in detail for LiV$_2$O$_4$ by present
authors.\cite{Yushankhai07}.  An analysis of inelastic neutron
scattering (INS) measurements\cite{Krimmel99,Lee01,Murani04} and
calculations of the dynamic spin susceptibility allowed us to suggest
the location of the paramagnetic spinel LiV$_2$O$_4$ close to a
magnetic instability. This was achieved by developing the RPA
theory of spin fluctuations based on {\it ab-initio} band structure
calculations and an on-site Coulomb interaction of 3$d$ electrons.
Close to the critical value of the interaction strength, low energy
spin fluctuations develop throughout a large shell in momentum space.
They may be mapped to an effective low energy paramagnon model which
describe low-temperature INS results\cite{Krimmel99,Lee01,Murani04}
accurately. From the comparison at $T\to$ 0, the parameters of the
model (peak energy, weight and extension in momentum space) are
fixed. Using the selfconsistent renormalization (SCR)
theory,\cite{Moriya73,Hasegawa74,Moriya85} which includes
mode-coupling of spin fluctuations,  the finite $T$ properties of INS
spectral shapes, uniform and staggered susceptibility, as well as NMR
relaxation rate, have been explained \cite{Yushankhai07,Yushankhai08,
Yushankhai2008}.  From this analysis we concluded that LiV$_2$O$_4$
can be regarded as a nearly antiferromagnetic (AFM) metal and its
unusual low-$T$ properties have to be related to a peculiar structure
of the paramagnetic ground state with strongly degenerate low-energy
(slow) AFM spin fluctuations.

In the present study, our main concern is to explain
the low-temperature, $T <$ 40K, electrical resistivity $\rho(T)$
measured on single crystals of LiV$_2$O$_4$ and reported by Takagi
{\it et al.}\cite{Takagi99} and Urano {\it et al.}\cite{Urano00}.
A Fermi-liquid behavior $\rho(T)\sim T^2$ for $T <$ 2K and
a more slow increase of $\rho(T)$ for higher temperatures were found.
Measurements\cite{Takagi99,Urano00} revealed a noticeable change in
physical properties of LiV$_2$O$_4$ for $T >$ 40K, including a
Curie-Weiss magnetic susceptibility $\chi(T)$ and a highly incoherent
transport, which is, however, beyond the scope of present theory.
In our approach we will use the effective low energy paramagnon
model for spin fluctuations whose parameters are completely fixed
by the comparison with INS. Only two more pheonomenological parameters
characterising the impurity and paramagnon scattering mechanism will
be needed.

From an analysis of the optical reflectivity and conductivity
measurements, J\"onsson {\it et al.}\cite{Jonsson07} and Irizawa {\it
et al.}\cite{Irizawa10} inferred that the conducting electron system
in LiV$_2$O$_4$ at ambient pressure is located close to a
correlation-driven insulating state. Under the applied external
pressure,\cite{Irizawa10} the system undergoes a metal-insulator
transition accompanied with a charge ordering and a structural
lattice distortion in the insulating phase. The observed complicated
phase transformation display properties different from those expected
from a first-order phase transition or a conventional metal-insulator
transition. We note that because of quarter-filling of the electronic
$t_{2g}$ bands in LiV$_2$O$_4$, electron correlations due to
inter-site Coulomb repulsion have to play an essential role in the
observed transition.\cite{Yushankhai05} Under such conditions, a
microscopic mechanism for the heavy quasiparticle formation on the
metallic side of the transition has to be clarified.  However, in
the present analysis which is concerned with resistivity under
ambient pressure the effect of inter-site Coulomb interaction and a
slowing-down of charge fluctuations  will not be included.

It was realized some time ago that the calculation of low-$T$ transport
properties in nearly AFM metals is a rather subtle issue.
\cite{Hlubina95,Rosch99,Rosch00} In clean systems with a peculiar AFM
ordering vector ${\bf Q}_{AFM}$ the quasiparticle scattering by
quantum-critical spin fluctuations is strongly anisotropic. The
strongest scattering occurs near the "hot spots" of the Fermi surface
(FS) connected by ${\bf Q}_{AFM}$ and the main contribution to the
electrical conductivity is due to quasiparticles from the "cold
region" of the FS. In that case, if the system is at some distance
from the AFM quantum critical point, the low-$T$ scattering rates are
proportional to $T^2$ and the Fermi-behavior $\rho(T)-\rho_{imp}\sim
T^2$ is realized. As was first pointed out by
Rosch,\cite{Rosch99,Rosch00} if a small amount of disorder is
present, an interplay of strongly anisotropic scattering due to
critical spin fluctuations and an isotropic impurity scattering may
complicate the picture producing  several different regimes for
$\rho(T)-\rho_{imp}\sim T^{\alpha}$  with the exponent between  1 $\leq \alpha \leq$ 2 in the low-$T$ region.

To describe temperature-dependent  electrical resistivity in
LiV$_2$O$_4$, we suggest that quasiparticles are scattered by AFM
spin fluctuations almost equally strongly over the FS and effects of
anisotropy are weak. This is related to a peculiar distribution of
dominant AFM spin fluctuations with ordering vectors ${\bf Q}_c$'s
forming a largely isotropic dense manifold in ${\bf k}$ space in this
compound.\cite{Yushankhai07} Low-temperature evolution of interacting
spin fluctuations in LiV$_2$O$_4$ can be successfully described
within the SCR formalism as presented in our previous
studies.\cite{Yushankhai08,Yushankhai2008} As explained in
Ref.\onlinecite{Moriya85},  the set of model parameters of the SCR
theory were obtained\cite{Yushankhai08} from neutron scattering
data\cite{Krimmel99,Lee01,Murani04} and used to
describe\cite{Yushankhai2008} the temperature and pressure evolution
of the spin-relaxation rate $1/T_1T$ observed in the NMR
measurement\cite{Fujiwara04} of the low-$T$ spin dynamics in
LiV$_2$O$_4$.  In this work, the theory is extended and applied  to
give an explanation of the low-$T$ electrical resistivity $\rho(T)$
in this compound.\cite{Takagi99,Urano00}

\section{Variational principle for ${\bf \rho(T)}$: general consideration.}

As follows from experimental
observations,\cite{Kondo97,Johnston00,Kondo99,Takagi99,Urano00} the concept
of the Fermi quasiparticles for charge carriers in the metallic
spinel  LiV$_2$O$_4$ is valid for sufficiently low temperatures,
$T<$ 30 K. In this regime, we assume that the dominant scattering
processes are given by low-energy AFM spin fluctuations and
impurities. In the linear response theory, in an applied electric
field $\bf E$ the quasiparticle distribution function $f_{\bf k}$ is
linearized around the equilibrium Fermi distribution $f^0_{\bf k}$
according to $f_{\bf k}=f^0_{\bf k}-\Phi_{\bf k}df^0_{\bf
k}/d\epsilon_{\bf k}$. The electronic transport can be found from the
Boltzmann equation
\begin{equation}
-e({\bf Ev_{k}})\frac{df^0_{\bf k}}{d\epsilon_{\bf
k}}=\sum\limits_{\bf k'}W_{\bf kk'}\Phi_{\bf k'}. \label{a1}
\end{equation}
The scattering operator $W_{\bf kk'}$ can be expressed through the total equilibrium transition
probability ${\mathcal P}_{\bf kk'}= {\mathcal P}_{\bf kk'}^{imp}+{\mathcal P}_{\bf kk'}^{sf}$ as
($k_B = \hbar =1$):
\begin{equation}
W_{\bf kk'}=\frac{1}{T}\left(\delta_{\bf kk'}\sum\limits_{\bf k''}{\mathcal P}_{\bf kk''} -
{\mathcal P}_{\bf kk'}\right), \label{a2}
\end{equation}
provided the spin fluctuations  are in thermal equilibrium, i.e.,
there is no drag effect.

For the elastic impurity scattering one has
\begin{equation}
{\mathcal P}_{\bf kk'}^{imp}=2\pi n_i|T_{\bf
kk'}|^2\delta(\epsilon_{\bf k}- \epsilon_{\bf k'})f^0_{\bf
k}(1-f^0_{\bf k'}). \label{a3}
\end{equation}
To a sufficiently good approximation, the $T$-matrix in Eq.(\ref{a3})
is frequently assumed  to be a constant $|T_{\bf kk'}|^2\approx
V_{imp}^2$ and $n_iV_{imp}^2$, where $n_i$ is the impurity density,
is regarded as a free parameter to be chosen so as to give a
realistic value of the measured residual resistivity $\rho_{imp}$. We
avoid this approximation and treat below matrix elements of
${\mathcal P}_{\bf kk'}^{imp}$ generally.

For the spin-fluctuation ($sf$) scattering one has\cite{Ueda75,Ueda77,Hlubina95,Rosch00}
\begin{equation}
{\mathcal P}_{\bf kk'}^{sf}=3J_{sf}^2f^0_{\bf k}(1-f^0_{\bf
k'})[n(\epsilon_{\bf k}- \epsilon_{\bf k'})+1] \mbox{Im}
\chi\left({\bf k}-{\bf k'}, \epsilon_{\bf k}- \epsilon_{\bf
k'}\right),\label{a4}
\end{equation}
where $n(\epsilon)$ is the Bose distribution function, $\chi\left({\bf q},\epsilon\right) $ is the
dynamical spin susceptibility describing the low-$T$ paramagnetic state of LiV$_2$O$_4$ and
$J_{sf}$ is an effective coupling constant which is the second free parameter. It is worth
emphasizing that in the present study the other parameters of the phenomenological SCR theory
determining the behavior of $\chi\left({\bf q},\epsilon\right)$ are considered to be known and
fixed from a fit to the data of inelastic neutron scattering measurement
\cite{Krimmel99,Lee01,Murani04} on LiV$_2$O$_4$, as discussed in Ref.\onlinecite{Yushankhai08}.

Following the standard notation\cite{Ziman60}, the Boltzmann equation (\ref{a1}) can be rewritten
in the form $X_{\bf k}=\sum\nolimits_{\bf k'}W_{\bf kk'}\Phi_{\bf k'}$. Then, the electrical
resistivity can be obtained by minimizing a functional\cite{Ziman60}
\begin{equation}
\rho[\Phi]=min \left[\frac{\langle \Phi,W\Phi \rangle}{|\langle\Phi,X(E=1)\rangle|^2}\right].
\label{a5}
\end{equation}
Here, $E=1$  means the unit electrical field and the scalar product of two functions $\Phi_{\bf k}$ and
$\Psi_{\bf k}$ is defined as $\langle \Phi,\Psi \rangle =
\sum\nolimits_{\bf k}\Phi_{\bf k}\Psi_{\bf k}$. In fact, in
Eq.(\ref{a5}) the ${\bf k}$-integration over the actual FS is
implied, which follows from the property of the scattering operator
$W_{\bf kk'}$ and the explicit form of $X_{\bf k}= e({\bf
Ev_{k}})(-df^0_{\bf k}/d\epsilon_{\bf k})$.

A way to search for a variational solution of Eq.(\ref{a5}) for the deviation function $\Phi_{\bf
k}$ is  to expand it in a set of the Fermi-surface harmonics (FSH) $\phi_L(\bf k)$:
\begin{equation}
\Phi_{\bf k}=\sum_L \eta_L \phi_L(\bf k), \label{a6}
\end{equation}
where $\eta_L$ are variational parameters and $L$ is a convenient composite label that includes
numbering of different sheets of the FS in  LiV$_2$O$_4$. The FSH's are
defined\cite{Allen76,Allen78} as polynomials of the Fermi-velocity Cartesian components $v_{\bf
k}^{\alpha}$. That is, for each integer $N\geq 0$ one has to construct $(N+1)(N+2)/2$ polynomials
$(v_{\bf k}^{x})^l(v_{\bf k}^{y})^m(v_{\bf k}^{z})^n$ with $l,m,n\geq 0$ and $l+m+n=N$, and
orthonormalize them on the actual FS, $\langle \phi_{L'},\phi_L \rangle =\delta_{L'L}$. The
resulting polynomials forming a complete
 set $\{\phi_L(\bf k)\}$  of basis functions are classified according to different irreducible
representations $\Gamma$ of the lattice symmetry point group.
In general, for a given $\Gamma$ there are subsets of different functions,
$\{\phi_{L}^r(\bf k)\}$, $\{\phi_L^s(\bf k)\}$, etc., which
transform according to the same  $\Gamma$. Then, for any
pair of partner functions, $\phi_{L'}^r(\bf k)$ and $\phi_{L''}^s(\bf k)$, belonging to different subsets,
but transforming  according to the same row of  $\Gamma$,  one has $\langle \phi_{L'}^r,W\phi_{L''}^s
\rangle\not=0$.   The other off-diagonal matrix elements of the scattering operator
$W_{\bf kk'}$, including those connecting different irreducible  representations, vanish by symmetry
arguments and, hence, the scattering operator has a block-diagonal form (see the discussion by
Allen\cite{Allen76} and references therein).

A minimum of $\rho[\Phi]$ is achieved in the class of odd functions, $\phi_L(-{\bf k})=-\phi_L({\bf
k})$; only these basis functions are included in the expansion (\ref{a6}).  Recalling the cubic
symmetry of the LiV$_2$O$_4$ lattice structure, we assume without loss of generality that the
applied electric field points in the ${\bf x}$ direction, which immediately distinguishes one of
the first-order FSH's: $\phi_{1x}({\bf k})=v_{{\bf k}}^{x}/\langle(v_{\bf k}^{x})^2\rangle^{1/2}$,
where $\langle(v_{\bf k}^{x})^2\rangle^{1/2}$ is for the root-mean-square on the Fermi surface.

A general strategy in describing the physical resistivity $\rho(T)$ as a solution of the
variational equation (\ref{a5}) in most of the metallic systems, including those with complicated
electronic band structure, is to truncate the expansion (\ref{a6}) by keeping in it only a few of FSH's.
 Following the common practice, one may start the analysis  with the
lowest, first-order variational solution, $\Phi_{\bf k}^{(0)}\sim
\phi_{1x}({\bf k})$, which is a fairly good approximation provided
the anisotropic effects of the quasiparticle scattering are weak. As
usual, here the anisotropy of the scattering operator $W_{\bf kk'}=
W_{\bf kk'}^{imp}+W_{\bf kk'}^{sf}$ means that the transition
probabilities depend not only on the mutual angle between the momenta
${\bf k}$ and ${\bf k'}$, but also on their position with respect to
the crystallographic axes. Anisotropic effects, as well as a
complexity of the actual FS, can be partially caught in the
calculations by keeping in the expansion (\ref{a6}) a selected number
of higher-order FSH's.

To go beyond the lowest-order solution for $\rho[\Phi]$ in the simplest manner, the following
approximate assumption can be made: the off-diagonal matrix elements $W_{LL'}=\langle
\phi_L,W\phi_{L'} \rangle$ are small compared to diagonal ones,  $W_{LL}$ and  $W_{L'L'}$. Then,
the variational solution to Eq.(\ref{a5}), being written in the familiar form\cite{Ziman60} as
 $\rho^{-1}=X_1^2\langle \phi_{1x},W^{-1}\phi_{1x} \rangle$, where $X_1=\langle
\phi_{1x},X\rangle$, can be expanded in a perturbation series\cite{Allen78}
\begin{equation}
\rho \approx \frac{1}{X_1^2}W_{1x,1x}\left[ 1 -
{\sum_L}'\frac{W_{1x,L}W_{L,1x}}{W_{1x,1x}W_{LL}}+
{\sum_{LL'}}'\frac{W_{1x,L}W_{LL'}W_{L',1x}}{W_{1x,1x}W_{LL}W_{L'L'}}-...
\right],\label{a7}
\end{equation}
where the primes on the sums means that the terms with $L,L'=1x$ and
$L=L'$ are excluded.

For $T\to$ 0, from Eqs.(\ref{a2}) and  (\ref{a4})  one has $W_{LL'}^{sf}\to$ 0, both for $L=L'$ and
$L\not=L'$, and the Eq.(\ref{a7}) reduces to
\begin{equation}
\rho(T\to 0)= \rho_{imp}\approx X_1^{-2}W_{1x,1x}^{imp}\zeta,
\label{b7}
\end{equation}
where the constant $\zeta$ stands for brackets in Eq.(\ref{a7}) with $W_{LL'}=W_{LL'}^{imp}$. Its
value is less than unity, 0 $< \zeta <$ 1, since the inclusion of higher-order terms leads a lower
estimate of the upper bound for $\rho$. The expression given by Eq.(\ref{b7}) approximates the
experimental value\cite{Takagi99,Urano00} of the residual resistivity $\rho_{imp}^{exp}\approx$
32 $\mu \Omega {\rm cm}$ in a low-$T$ fit procedure.

For the further purposes, we note that relations between $W_{1x,1x}^{imp}$ and the other diagonal
matrix elements $W_{LL}^{imp}$ cannot be generally established. In particular, a strong inequality
$W_{LL}^{imp}\gg W_{1x,1x}^{imp}$ for some $L\not=1x$ is not excluded, which does not invalidate
our previous assertions. Actually,  irrespective of a relation between $W_{1x,1x}^{imp}$ and
$W_{LL}^{imp}$,  the series expansion (\ref{a7}) starts with the matrix element $W_{1x,1x}^{imp}$
due to the requirement that the variational solution for $\rho$ is given by the diagonal
matrix element of the inverse scattering operator $W^{-1}$ between the same first-order FSH, i.e.,
$\rho^{-1}=X_1^2\langle \phi_{1x},W^{-1}\phi_{1x} \rangle$.

\section{Electron scattering by spin fluctuations in LiV$_2$O$_4$.}
An open question is: Whether one may rely an analysis of the physical
resistivity  on the series expansion (\ref{a7}) for $T >$ 0? Since
the impurity scattering is thought to be highly isotropic, one of the
underlying assumptions that $W_{LL'}^{imp}/W_{LL}^{imp}\ll$ 1 for
$L'\not= L$, has to be fulfilled. Here, the appearance of some
off-diagonal matrix elements $W_{LL'}^{imp}$ can be explained mostly
due to a complex character of the multi-sheet FS in LiV$_2$O$_4$.
Below we examine how  properties of AFM spin fluctuations are related
to those of the spin-fluctuation scattering operator $W_{{\bf
kk'}}^{sf}$ in LiV$_2$O$_4$, and show that the smallness of the
off-diagonal elements $W_{LL'}^{sf}$ with respect to diagonal ones,
$W_{LL}^{sf}$ and $W_{L'L'}^{sf}$, seems to be a plausible assumption
as well.

From Eqs.(\ref{a2}) and (\ref{a4}), any diagonal or off-diagonal
matrix element $W_{LL'}^{sf}$ allowed by symmetry arguments can be
written as follows
\begin{equation}
W_{LL'}^{sf}=\frac{1}{2T}\sum_{\bf kk'}\left[\phi_L({\bf k})-\phi_L({\bf k'})\right]{\mathcal
P}_{\bf kk'}^{sf}\left[\phi_{L'}({\bf k})-\phi_{L'} ({\bf k'})\right]. \label{a8}
\end{equation}

The use of the definition (\ref{a4}) for ${\mathcal P}_{\bf
kk'}^{sf}$ leads to
\begin{equation}
W_{LL'}^{sf} = \frac{1}{2T}\left(\frac{1}{2\pi}\right)^6\int
d\epsilon \oint \frac{d^2k}{v_{\bf k}} \oint \frac{d^2k'}{v_{\bf k'}}
\left[\phi_L({\bf k})-\phi_L({\bf k'})\right] {\mathcal P}^{sf}({\bf
k'}-{\bf k},\epsilon) \left[\phi_{L'}({\bf k})-\phi_{L'}({\bf
k'})\right], \label{a9}
\end{equation}
where the standard replacement $\sum_{\bf k}\to \int d\epsilon
\oint_{\epsilon}d^2k/[(2\pi)^3v_{\bf k}]$ for a unit volume  together with the relation $f^0(\epsilon_{\bf
k})[1-f^0(\epsilon_{\bf k'})][ n(\epsilon_{\bf k}-\epsilon_{\bf k'})+
1 ]= [f^0(\epsilon_{\bf k'})-f^0(\epsilon_{\bf k})]n(\epsilon_{\bf
k'}-\epsilon_{\bf k})[n(\epsilon_{\bf k'}-\epsilon_{\bf k})+1]$, and the
approximation  $[f^0(\epsilon)-f^0(\epsilon')]\approx
(\epsilon'-\epsilon)(-df^{0}/d\epsilon)$ are used; two-dimensional
integrations over $k$ and $k'$ are restricted to the FS. Then the
kernel ${\mathcal P}({\bf k'}-{\bf k},\epsilon)$ in Eq.(\ref{a9})
takes the form
\begin{equation}
{\mathcal P}^{sf}({\bf k'}-{\bf k},\epsilon) = 3J_{sf}^2  \epsilon
n(\epsilon) [n(\epsilon)+1] \mbox{Im} \chi\left({\bf k'}-{\bf k},
\epsilon \right). \label{a10}
\end{equation}
This is the well known form of conduction electron scattering
from spin fluctuations. The latter are enhanced by the nearly
critical Coulomb interaction of 3d electrons, which
leads\cite{Yushankhai07} to a paramagnon expression for
$\chi\left({\bf k'}-{\bf k},\epsilon \right)$  whose parameters are
fixed from INS results at $T\to$ 0.  At low temperatures, the
low-energy ($\epsilon \sim$ 1 meV) dynamic spin susceptibility
$\chi\left({\bf q'}, \epsilon \right)$ in LiV$_2$O$_4$ shows maxima
around the critical wave vectors ${\bf q' }={\bf Q}_c$ forming a
rather dense manifold $\{{\bf Q}_c\}$ in ${\bf k}$
space.\cite{Yushankhai07}  To take into account explicitly all
scattering processes due to dominant spin fluctuations, it is helpful
to make in Eq.(\ref{a10}) the following substitution
\begin{equation}
\mbox{Im} \chi\left({\bf k'}-{\bf k}, \epsilon \right)\simeq \sum_{\{{\bf Q}_c\}}\int
\frac{d^3q}{(2\pi)^3}\delta({{\bf k'}-{\bf k}-{\bf Q}_c-{\bf q}})\mbox{Im} \chi\left({\bf Q}_c+{\bf
q}, \epsilon \right), \label{a11}
\end{equation}
which ascribes  particular weights $\mbox{Im} \chi\left({\bf
Q}_c+{\bf q}, \epsilon \right)$ to the quasiparticle scattering
processes whose wave vectors, ${\bf k'}$ and ${\bf k}$, at the Fermi
surface satisfy the relation ${\bf k'}-{\bf k}={\bf Q}_c+{\bf q}$. In
Eq.(\ref{a11}), the summation is over the entire set $\{{\bf Q}_c\}$
of the critical wave vectors and their neighborhoods, $|{\bf q}|\ll
|{\bf Q}_c|$. In total, this involves a broad region in  ${\bf k}$
space, where the dominant AFM spin fluctuations are distributed, and
$\mbox{Im} \chi\left({\bf Q}_c+{\bf q}, \epsilon \right)$ in
Eq.(\ref{a11}) does not much depend on a direction of ${\bf Q}_c$.
The resulting distribution differs strongly from that occurring at
low $T$ in most of nearly AFM metals where the low energy
susceptibility $\chi\left({\bf q}, \epsilon \right)$ is usually
peaked around a discrete ordering wave vector ${\bf Q}_{AFM}$.

The use of the above arguments  allows us to write down the matrix element $W_{LL'}^{sf}$ in a
factorized form
\begin{equation}
W_{LL'}^{sf}\approx C_{LL'}{\mathcal F}(T), \label{a12}
\end{equation}
where
\begin{equation}
C_{LL'} = \left(\frac{1}{2\pi}\right)^6\oint \frac{d^2k}{v_{\bf k}}
\oint \frac{d^2k'}{v_{\bf k'}} \left[\phi_L({\bf k})-\phi_L({\bf
k'})\right] M_{{\bf k}{\bf k'}}^{sf}\left[\phi_{L'}({\bf
k})-\phi_{L'}({\bf k'})\right], \label{a13}
\end{equation}

\begin{equation}
{\mathcal F}(T)=\frac{1}{T}\int\limits_0^{\infty}d\epsilon \int \frac{d^3q}{(2\pi)^3}\epsilon
n(\epsilon)[n(\epsilon)+1]\mbox{Im} \chi\left({\bf Q}_c+{\bf q},\epsilon \right), \label{a14}
\end{equation}
and the matrix $M_{{\bf k}{\bf k'}}^{sf}$ is defined as $M_{{\bf k}{\bf k'}}^{sf}\simeq
3J_{sf}^2\sum_{\{{\bf Q}_c\}}\delta({{\bf k'}-{\bf k}-{\bf Q}_c})$. The matrix is invariant under
simultaneous operations of the lattice point group on both ${\bf k}$ and ${\bf k'}$, since the
manifold ${\{{\bf Q}_c\}}$ is an invariant as well.  The wave vectors ${\bf Q}_c$ are along all high
symmetry directions in the Brillouin zone (BZ) and their end points are lying on a closed surface
of a mean radius $|{\bf Q}_c|\sim 0.6 \AA^{-1}$ which is referred to as the "critical"
surface.\cite{Yushankhai07}  The factorization of  matrix elements introduced by  Eq.(\ref{a12}) implies that the
quasiparticle scattering by spin fluctuations with different ${\bf Q}_c$  at the "critical" surface provide  nearly identical
contributions and, therefore, only one representative  wave vector
${\bf Q}_c$ appears in  Eq.(\ref{a14}).

As discussed in Ref.\onlinecite{Yushankhai07}, the high directional degeneracy of ${\bf Q}_c$'s
results from  complexity of the electronic band structure and the geometrical frustration of the
pyrochlore lattice structure of LiV$_2$O$_4$. In this respect, the low-$T$ spin-fluctuation
scattering mechanism\cite{Hlubina95,Rosch99,Rosch00} operating with a peculiar ordering wave vector
${\bf Q}_{AFM}$ differs from that occurring in the paramagnetic spinel LiV$_2$O$_4$. In the former
case, the quasiparticle scattering  is a strongly anisotropic one, leading, for instance, to "hot
spots" at the Fermi surface. Instead,  the above analysis suggests that the quasiparticle
scattering  by spin fluctuations in LiV$_2$O$_4$ is largely isotropic one. In that case, the
diagonal matrix element $W_{LL}^{sf}$ prevail over the off-diagonal ones. This fact and similar
arguments mentioned above for the impurity scattering justify the applicability of the perturbation
series expansion, Eq.(\ref{a7}), for $\rho(T)$.
%
\begin{figure}
\includegraphics[width=8cm]{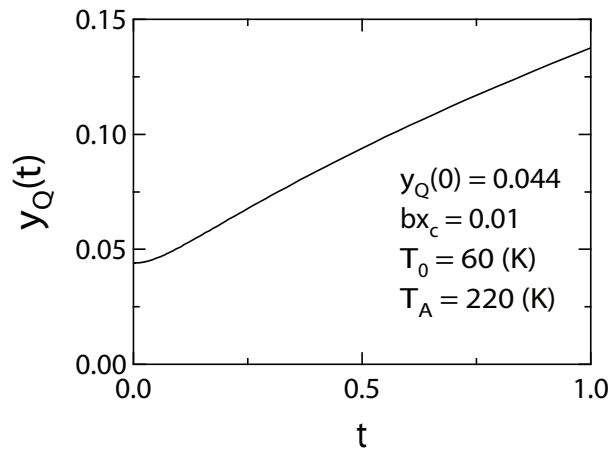}
\caption{The solution of the basic equation of the phenomenological SCR theory  for the reduced inverse
static spin susceptibility  $y_Q(t)$ as a function of the reduced temperature $t=T/T_0$;
phenomenological parameters required to obtain $y_Q(t)$ in LiV$_2$O$_4$ are  given in the text.}
\label{fig:Fig1}
\end{figure}
%

\section{Calculation of $\rho(T)$ based on the SCR theory of spin fluctuations in LiV$_2$O$_4$.}
First, based on the expansion (\ref{a7}), the resistivity $\rho(T)$
can be expressed as follows
\begin{equation}
\rho(T)-\rho_{imp}=\rho^{(1)}_{sf}(T) + \Delta \rho(T), \label{a15}
\end{equation}
where the spin-fluctuation contribution $\rho_{sf}^{(1)}(T)=
X_1^{-2}W_{1x,1x}^{sf}$ represents the lowest-order solution and
$\Delta \rho(T)$ is a correction  due to higher-order terms in
Eq.(\ref{a7}). Their variations with $T$ can be found by calculating
the function ${\mathcal F}(T)$, Eq.(\ref{a14}), entering the matrix
elements of the spin-fluctuation scattering operator, Eq.(\ref{a12}).
Although the subsequent derivation of an explicit form of ${\mathcal
F}(T)$ has much in common with earlier
studies\cite{Ueda77,Kondo02} of nearly AFM metals, essential
features specific to LiV$_2$O$_4$ have to be emphasized. In
particular, within the SCR theory of spin fluctuations the imaginary
part of the dynamic spin susceptibility can be parametrized as
follows\cite{Yushankhai08}
\begin{equation}
\mbox{Im}\chi\left({\bf Q}_c+{\bf q}, \epsilon;
T\right)=\frac{1}{4\pi T_AT_0}\frac{\epsilon}{\left[
y_Q\left(T\right) + \left(q^{||}/q_B\right)^2  + b\left({\bf
q}^{\bot}/q_B\right)^2\right]^2 + \left(\epsilon/2\pi T_0\right)^2}.
\label{a16}
\end{equation}
Here, for a given ${\bf Q}_c$, $q^{||}$ and ${\bf q}^{\bot}$ are the
components of ${\bf q}$ parallel and perpendicular to ${\bf Q}_c$,
respectively; $q_B$ is the effective radius of the BZ boundary given
in terms of a primitive cell volume $v_0$ as
$q_B=\left(6\pi^2/v_0\right)^{1/3}$. The parameters $T_A \simeq$ 220K
and $T_0 \simeq$ 60K characterize the widths of the momentum and
energy distributions of spin fluctuations, respectively; a small
parameter $b$ takes care about a strong anisotropy of the
distribution in ${\bf k}$ space.
 Next, the reduced inverse susceptibility at ${\bf Q}_c$ is defined as $y_Q\left(T\right)=
 [2T_A\chi\left(Q_c; T\right)]^{-1}$ assuming a "spherical" approximation, i.e.
 $\chi\left({\bf Q}_c; T\right)$ does not depend on a direction of ${\bf Q}_c$.

With  the insertion of Eq.(\ref{a16}) into Eq.(\ref{a14}), we get
\begin{equation}
{\mathcal F}(T)=\frac{2\pi T_0}{T_A}\int \frac{d^3q}{(2\pi)^3} \int\limits_0^{\infty}d\lambda
n(\lambda)[n(\lambda)+1] \frac{\lambda^2}{\lambda^2+(2\pi u_{\bf q})^2}= \frac{2\pi T_0}{T_A}\int
\frac{d^3q}{(2\pi)^3} I(u_{\bf q}). \label{a17}
\end{equation}
Here, the last equality defines $I(u_{\bf q})$, where
\begin{equation}
u_{\bf q}(t)=\frac{y_Q\left(t\right) + \left(q^{||}/q_B\right)^2  + b\left({\bf
q}^{\bot}/q_B\right)^2}{t}, \label{a18}
\end{equation}
and $t=T/T_0$ is the reduced temperature. At the next step, the function $I(u_{\bf q})$ can be
expressed as
\begin{equation}
 I(u_{\bf q})=\frac{u_{\bf q}}{2}\left[\psi'(u_{\bf q}) - \frac{1}{u_{\bf q}} - \frac{1}{2u_{\bf q}^2} \right], \label{a19}
\end{equation}
where $\psi(u)$ is the digamma function and $\psi'(u)= d\psi(u)/du$.

The integration over ${\bf q}$ in Eq.(\ref{a17}) are performed by using the same prescriptions as
in Ref.\onlinecite{Yushankhai08}, which yields  $(2\pi)^{-3}\int d^3q I(u_{\bf q})= c {\bar
f}(T/T_0)$, where $c$ is the known dimensionless factor, $c\approx 4$, and\cite{Kondo02}
\begin{eqnarray}
{\bar
f}(T/T_0)=\frac{t}{bx_c}&&\hspace*{-4mm}\int\limits_0^{z_c}dz\left[-\frac{bx_c}{t}-\frac{1}{2}\ln
\left(\frac{y_Q(t)+ z^2  + bx_c}{y_Q(t) + z^2}\right) \right.+\nonumber\\
&&\frac{y_Q(t) + z^2  + bx_c}{t} \psi \left(\frac{y_Q(t)+ z^2 + bx_c}{t}\right) -\frac{y_Q(t) +
z^2}{t}\psi \left(\frac{y_Q(t) + z^2}{t}\right) - \nonumber\\
&&\left.\ln \Gamma \left(\frac{y_Q(t) + z^2  + bx_c}{t}\right)+ \ln \Gamma \left(\frac{y_Q(t) +
z^2}{t} \right)\right], \label{a20}
\end{eqnarray}
where $\Gamma(u)$ is the gamma function, the cutoff $z_c\simeq$1/2 and the remaining parameter
$bx_c\simeq10^{-2}$.

For a given $y_Q(t)$, the expression (\ref{a20}), as a function of temperature, can be calculated
numerically. This determines, according to  Eqs.(\ref{a12}),(\ref{a17})-(\ref{a20}), an evolution
with $T$ of any non-vanishing matrix element $W_{LL'}^{sf}$ in the whole range $T <$ 40K, where the
SCR theory for the AFM spin fluctuations in LiV$_2$O$_4$ is proved to be valid\cite{Yushankhai08}.
Here we utilize the known functional form for $y_Q(t)$, Fig.1,  obtained by solving the basic
equation of the SCR theory developed in Ref.\onlinecite{Yushankhai08} to explain results of
inelastic neutron scattering measurements on LiV$_2$O$_4$. The solution shows that $y_Q(t)$ is a
monotonically increasing function of temperature; the limiting value $y_Q(t\to 0)$, was found to be
$y_Q(0)=$0.044. The use of the energy scale, $T^{\ast}=2\pi T_0 y_Q(0)\approx$ 16K, which is the
relaxation rate of the low-energy spin fluctuations,\cite{Yushankhai08}  is helpful in recognizing
two regimes with different power-law behavior of $y_Q(t)$. Actually, for $T\ll T^{\ast}$, one
obtains $[y_Q(t)-y_Q(0)]/y_Q(0)\ll 1$, which leads to the quadratic behavior of ${\bar
f}(T/T_0)\sim T^2$. For $T^{\ast}<T$, a smooth, nearly linear, $t$-dependence of $y_Q(t)$ results
in a peculiar monotonic temperature increase of ${\bar
f}(T/T_0)$, as indicated below. \\

 First, we analyze the lowest-order approximation to the low temperature  ($T\ll T^{\ast}$) resistivity,
\begin{equation}
\rho(T) \approx  \rho_{imp}+\rho^{(1)}_{sf}(T), \label{b20}
\end{equation}
where
\begin{equation}
\rho^{(1)}_{sf}(T)= {\mathcal A}^{(1)}_{sf}{\bar f}(T/T_0),
\label{c20}
\end{equation}
and compare its  $T$-dependence with that of the observed\cite{Takagi99} experimental resistivity. Here
${\mathcal A}^{(1)}_{sf}=(2\pi c T_0/T_A)X_1^{-2}C_{1x,1x}$ is an adjustable parameter (together
with $\rho_{imp}$) in a low-$T$ fit procedure using the calculated ${\bar f}(T/T_0)$ shown in
Fig.2.

One may see that the function ${\bar f}(T/T_0)$ nearly precisely
follows the quadratic dependence, ${\bar f}(T/T_0)=c_1T^2$  with $c_1
=$ 0.0033, for $T <$ 2K $\ll T^{\ast}$, where the Fermi-liquid
behavior $[\rho^{exp}(T)-\rho_{imp}]=AT^2$ in LiV$_2$O$_4$ was
reported.\cite{Takagi99,Urano00} From the low-$T$ fit procedure, as
shown in Fig.3, the parameter ${\mathcal A}^{(1)}_{sf}$ was found to
be ${\mathcal A}^{(1)}_{sf}=$ 666.7  $\mu \Omega cm$, which corresponds
to the observed coefficient of the $T^2$ term, $A=$ 2.2 $\mu \Omega
cm/K^2$.

With increasing $T$ and starting from  $T\approx$ 2K, both the calculated $\rho^{(1)}_{sf}(T)$ and
the measured resistivity $[\rho^{exp}(T)-\rho_{imp}]$ show gradual deviations from the $T^2$
behavior, however, with somewhat different rates. Specifically, starting from $T\approx$ 2K one
obtains the growing discrepancy $[\rho^{exp}(T)-\rho_{imp}]-\rho^{(1)}_{sf}(T)= \Delta\rho(T)<$ 0,
which indicate that the higher-order corrections involved in $\Delta\rho(T)$ cannot be longer
neglected. Remarkably, a negative correction, $\Delta \rho(T)<0$, is consistent with the
variational principle requiring that an extension of the involved basis functions should lead to an
improved upper bound on $\rho(T)$.
%
\begin{figure}
\includegraphics[width=8cm]{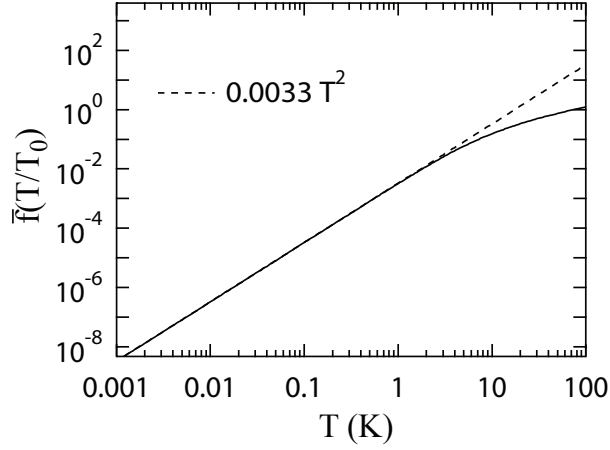}
\caption{$T$-dependence of ${\bar f}(T/T_0)$ defined by Eq.(\ref{a20}) and
calculated with the use of the reduced inverse susceptibility $y_Q(T/T_0)$ shown in Fig.1.}
\label{fig:Fig2}
\end{figure}
%
Actually, let us consider the lowest-order correction $\Delta
\rho^{(2)}_{sf}(T)$ which can be written from Eq.(\ref{a7}) as
\begin{equation}
\Delta \rho^{(2)}_{sf}(T)= -
\frac{1}{X_1^2}{\sum_L}'\left[\frac{(W_{1x,L}^{sf}+W_{1x,L}^{imp})^2}{W_{LL}^{imp}+W_{LL}^{sf}}
- \frac{(W_{1x,L}^{imp})^2}{W_{LL}^{imp}}\right],  \label{a21}
\end{equation}
where  the second term in brackets, being already involved in
$\rho_{imp}$, is now subtracted to ensure that $\Delta
\rho^{(2)}_{sf}(T\to 0)=$ 0. Note that a $T$-dependence in the
right-hand side of Eq.(\ref{a21}) is entirely due to
$W_{LL'}^{sf}=C_{LL'}{\mathcal F}(T)$, both for $L'=L$ and
$L'\not=L$. For sufficiently low $T$,  a denominator in the
right-hand side of Eq.(\ref{a21}) can be approximated assuming that
$C_{LL}{\mathcal F}(T)\ll W_{LL}^{imp}$ and ${\mathcal F}(T)\sim
T^2$, which immediately leads to  small leading correction, $\Delta
\rho^{(2)}_{sf}(T)\approx aT^2-|b|T^4 + O(T^6)$. Here, the first
quadratic term can be adopted by changing slightly a value of the
adjustable parameter ${\mathcal A}^{(1)}_{sf}$, while the next term,
$-|b|T^4$, provides the required negative correction to the
first-order result $\rho^{(1)}_{sf}(T)$. \\

An extension of the above analysis to higher temperature  could be possible if
one establishes reliable relations between numerous matrix elements
involved in Eq.(\ref{a7}). The following plausible assumption can be
made based on the fact that ${\mathcal F}(T)=(2\pi c T_0/T_A){\bar
f}(T/T_0$) is a rapidly growing function of $T$, Fig.2. For instance,
${\mathcal F}(T\sim 30\mbox{K})$/${\mathcal F}(T\sim 1\mbox{K})\sim
10^2$. We suggest that the limit, $C_{LL}{\mathcal F}(T)\gg
W_{LL}^{imp}$, can be achieved at $T\lesssim$ 40K, i.e near the
border where the SCR theory of spin fluctuations in LiV$_2$O$_4$ is
still valid. With this assumption, one obtains, for instance, the
following estimate for the second order correction, $\Delta
\rho^{(2)}_{sf}(T)\approx
-[X_1^{-2}{\sum_L}'(C_{1x,L})^2/C_{LL}]{\mathcal F}(T)+const$. Then,
by doing in the same manner and after collecting  all leading terms
in the expansion (\ref{a7}),  the physical resistivity in
LiV$_2$O$_4$ for 2K$\ll T \lesssim$ 40K can be approximated by the
following simple functional form
\begin{equation}
\rho(T)\approx \rho_{imp}+{\mathcal B}+{\mathcal A}_{sf}{\bar f}(T/T_0), \label{a22}
\end{equation}
where
\begin{equation}
{\mathcal A}_{sf}=\frac{2\pi c T_0}{T_A}\frac{1}{X_1^2}\left[
C_{1x,1x} - {\sum_L}'\frac{(C_{1x,L})^2}{C_{LL}} +
{\sum_{LL'}}'\frac{C_{1x,L}C_{LL'}C_{L',1x}}{C_{LL}C_{L'L'}}-...
\right].\label{a23}
\end{equation}
\begin{equation}
{\mathcal B}=\frac{1}{X_1^2} {\sum_L}'\left\{ W_{LL}^{imp}\left(\frac{C_{1x,L}}{C_{LL}}\right)^2 +
W_{1x,L}^{imp}\left[1-2\frac{C_{1x,L}}{C_{LL}}\right]-...\right\}\label{b23}
\end{equation}

By noting a close similarity  between Eq.(\ref{a22}) and
Eq.(\ref{b20}), it is worth emphasizing completely different
$T$-dependence of ${\bar f}(T/T_0)$ in the low- and a
high-temperature regimes. Moreover, the factor ${\mathcal A}_{sf}$ is
subjected to a special constraint with respect to ${\mathcal
A}_{sf}^{(1)}$. Actually, for $T < $ 2K $\ll T^{\ast}$ the
first-order term in the series (\ref{a23}) is only needed, ${\mathcal
A}_{sf}^{(1)} \simeq (2\pi c T_0/T_A) X_1^{-2} C_{1x,1x}$, while for
2K $\ll T \lesssim$ 40K, the factor $A_{sf}$ is given by the full
series (\ref{a23}) and, hence, ${\mathcal A}_{sf} < {\mathcal
A}_{sf}^{(1)}$ is required. An estimate for a shift ${\mathcal B}$ in
Eq.(\ref{a22}), which is the second adjustable parameter in a
high-$T$ fit procedure, is discussed bellow.
%
\begin{figure}
\includegraphics[width=8cm]{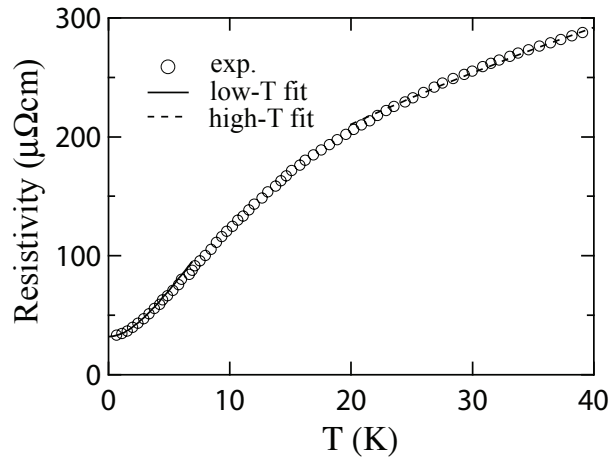}
\caption{Theoretical  fits to the experimental data (Ref.\onlinecite{Takagi99}, open circles) for the
 electrical resistivity in LiV$_2$O$_4$ in
 different temperature regions.
 In the low-$T$ limit, the resistivity is described by
 Eqs.(\ref{b20}),(\ref{c20})
 with $\rho_{imp}=32(\mu \Omega {\rm cm})$ and
 $A_{sf}^{(1)}=666.7(\mu \Omega {\rm cm})$. In the high
 temperature region, 20 K $<T<$ 40 K, where the SCR theory is
 still valid, the resistivity is  approximated by Eq.(\ref{a22})
 with $\rho_{imp}+{\mathcal B}=117(\mu \Omega {\rm cm})$ and
 $A_{sf}=280(\mu \Omega {\rm cm})$.}
\label{fig:Fig3}
\end{figure}
%
In Fig.3, the physical resistivity $[\rho^{exp}(T)-\rho_{imp}]$ is compared
with the predicted behavior, Eq.(\ref{a22}), for $T>T^{\ast}(\approx$
16K). A  satisfactory coincidence between the experimental data and
calculated results is achieved for 20K $< T <$ 40K with two fit
parameters ${\mathcal A}_{sf}= 280 \mu \Omega cm$ and ${\mathcal B}=
85 \mu \Omega cm$. While the expected constraint, ${\mathcal A}_{sf}
< {\mathcal A}_{sf}^{(1)}$ is fulfilled, the obtained large value of
${\mathcal B}\approx 3\rho_{imp}$ means that the Matthiessen
rule\cite{Ziman60} is severely violated. A possible mechanism for
this effect is the following. Recalling the estimate, Eq.(\ref{b23})
for ${\mathcal B}$, together with relations $\rho_{imp} \sim
W_{1x,1x}^{imp}$ and $\left(C_{1x,L}/C_{LL}\right)^2\ll$ 1, we
suggest that for some $L\not= 1x$ one has $W_{LL}^{imp}\gg
W_{1x,1x}^{imp}$, which explains why an estimate ${\mathcal B}\sim
\rho_{imp}$ is feasible.

So far the special attention has been paid to two limiting regimes of
low- and comparatively high-temperatures ($T <$ 40K), where the
series expansion, Eq.(\ref{a7}), for $\rho(T)$ reduces to very
similar forms, Eqs. (\ref{b20}) and (\ref{a22}), requiring two
adjustable parameters for a fit procedure in each regimes . We insist
that Eq.(\ref{a7}) should provide the interpolation $T$-dependent
function between the low- and high-$T$ limits as well. However, for
intermediate temperatures, one has $C_{LL}{\mathcal F}(T)\sim
W_{LL}^{imp}$, and the corresponding fit procedure, though being
possible, would require a larger number of adjustable parameters.
This could hardly give more insight into the problem under
consideration and therefore we omit such a procedure in our
discussion.

\section{Conclusion}
We have calculated the electrical resistivity $\rho(T)$ in the  paramagnetic metallic spinel
LiV$_2$O$_4$ treated as a nearly AFM Fermi-liquid for temperatures $T <$ 40K . Impurities and
strongly degenerate temperature-induced low-energy AFM spin fluctuations were supposed to provide
two main sources of the quasiparticle scattering and the resistivity. The self-consistent
renormalization theory developed earlier was applied to derive explicitly the temperature-dependent
matrix elements of the spin fluctuation scattering operator. The absence of hot spots of the Fermi
surface and a largely isotropic character of the quasiparticle scattering was deduced from a
peculiar, nearly spherical, shape of the spin-fluctuation distribution in the momentum space for
the paramagnetic ground state in LiV$_2$O$_4$. Comparatively weak anisotropic effects were assumed
to originate mainly from  a complex many-sheet structure of the Fermi surface in this compound. The
assumption allowed us to use the variational solution for the Boltzmann equation in the form of a
perturbative series expansion for $\rho(T)$. Our theory remains to be a phenomenological one since
unknown model parameters were found from the best overall fit to the temperature-dependent
$\rho^{exp}(T)$ measured on a single crystal of LiV$_2$O$_4$.

The resulting theoretical expression for $\rho(T)$ was shown to take
very similar simple forms in two limiting regimes for spin
fluctuations, which describe successfully experimental results for
$\rho(T)$ with a minimal set of two adjustable parameters in each
regime. These include the low temperatures, $T\ll T^{\ast}$ (where
$T^{\ast}\approx$ 16K is the characteristic energy scale of spin
fluctuations), and somewhat higher temperatures, $T^{\ast} < T <$
40K, respectively.

For $T >$ 40K the SCR theory of AFM spin fluctuations in LiV$_2$O$_4$ is no longer valid. As
discussed in Ref.\onlinecite{Yushankhai08}, and evidenced from
experiment\cite{Takagi99,Urano00,Krimmel99,Lee01,Murani04}, with increasing $T$ the AFM
fluctuations at $|{\bf q}|\simeq Q_c$ are suppressed and no more distinguished from those at other
wave vectors in the BZ; the system enters a spin localized regime compatible with the Curie-Weiss
behavior of $\chi({\bf q}=0)$ observed in LiV$_2$O$_4$ for $T > 50$ K. An explanation of incoherent
transport properties in this regime remains to be a challenging problem.

\section{Acknowledgments}
One of the authors (V.Yu.) acknowledges partial financial support from the Heisenberg-Landau
program.




\end{document}